  \documentclass[aps,
 pra,%
 amsmath,amssymb,
 showpacs,longbibliography,   
 reprint,%
]{revtex4-1}

 \usepackage{dcolumn}
 \usepackage{amsmath}
\usepackage{bm}

\begin{document}


%
 \newcommand{\jjlabel}[1]{ }
%
%
%

\title{Refutation of [Chyba and Hand, Phys. Rev. Applied 6, 014017 (2016)]: No Electric Power can be Generated from Earth's Rotation through its Own Magnetic Field}
\author{J. Jeener}
\email{JJEENER@ULB.AC.BE}
\affiliation{Universit\'{e} Libre de Bruxelles (CP-223), B-1050
Brussels, Belgium}
\date{\today}
\begin{abstract} 
In a recent article [Phys. Rev. Applied $\bm{6}$, 014017 (2016)], Chyba and Hand propose a new scheme to generate electric power continuously at the expense of Earth's kinetic energy of rotation, by using an appropriately shaped cylindrical shell of a well chosen conducting ferrite, rigidly attached to the Earth. No experimental confirmation is reported for the new prediction. 
In the present Refutation, I first use today's standard electromagnetism and essentially the same model as Chyba and Hand to show in a very simple way that no device of the proposed type  can produce continuous electric power, whatever its configuration or size, in agreement with widespread expectation. 
Next, I show that the prediction of nonzero continuous power by Chyba and Hand results from a confusion of frames of reference at a critical step of their derivation.
When the confusion is clarified, the prediction becomes exactly zero and the article under discussion appears as pointless. 
At the end, I comment about the persistent invocation by Chyba and Hand of the misleading legacy notion that quasi-static magnetic fields have an intrinsic velocity, and 
other questionable concepts.
 \end{abstract} 
\maketitle
\section{Introduction}
In a recent article \cite{Chyba:2016fk}, Chyba and Hand (C\&H) propose a new scheme to generate electric power continuously at the expense of Earth's kinetic energy of rotation, by using an appropriately shaped cylindrical shell of a well chosen conducting ferrite, rigidly attached to the Earth, interacting with Earth's own magnetic field.  Special attention was drawn on this important unexpected prediction in a concomitant summary published in Physics \cite{Lindley:2016fk}. No experimental confirmation of the effect was reported. 

In Sec.~\ref{JJderiv} of the present Refutation 
(submitted to Phys. Rev. Applied as a Comment)
I  give a very simple proof that no device of the type proposed by C\&H can generate {\em continuous} electric power, whatever its configuration or size . My derivation uses today's conventional electromagnetic theory 
 \footnote{This means Maxwell's equations, special relativity (in the weakly relativistic approximation whenever convenient), Lorentz force equation (\ref{geoc06}) with $\bm{B}$, $\bm{E}$, $\bm{F}$, and the velocity $\bm{v}$ of the charge $q$ measured in the same inertial frame, and the usual phenomenological relations: Ohm's law ,$\cdots$, as discussed in detail by e.g. Landau and Lifshitz \cite{Landau:1984fk, Landau:1971fk}, Reitz and Milford \cite{Reitz:1960fk}, or Jackson \cite{Jackson:1962fk}.}
%
and essentially the same model as C\&H: constant sources of the actual geomagnetic field rotating together with Earth, passive device fixed to Earth, however 
 with ``{\em steady state}'' explicitly implying that vector fields like $\bm{B}$ have 
 settled to time independent values
  {\em as seen by Earth bound observers} (see Appendix A).

Next, in Sec.~\ref{C+H1}, I show that the prediction of nonzero electric power production in Secs. IV to IX of C\&H is the consequence of a confusion of properties of frames of reference between Eq.~C\&H(25) and Eqs.~C\&H(26-27). When the confusion is clarified, the prediction becomes exactly zero, in agreement with 
widespread expectation and with 
my own general result.

Two mutually incompatible versions of non-relativistic electromagnetic theory are used concurrently in C\&H's article: (a) quantitative calculations (display equations and their derivation) follow the standard theory that is accepted since about one century 
\cite{Note1}, albeit 
with unjustified replacement of the actual Earth's 
magnetic field by its axisymmetric component,
 and (b) qualitative ``intuitive'' predictions pervade the discussion, based on C\&H's own interpretation of long abandoned assumptions like the existence of an intrinsic velocity of quasi-static magnetic fields and related concepts.
%
%
%
%
These deviations from today's conventional electromagnetic theory are further discussed in Sec.~\ref{questionable}, together with other questionable aspect of C\&H's article.
\section{Proof that no power production is possible by Chyba and Hand's type of device}
\jjlabel{JJderiv}\label{JJderiv}
In Ref.~\onlinecite{Chyba:2016fk}, and in the present comment, irrelevant 
complications are avoided by neglecting the acceleration of Earth's center in its orbit, so that convenient inertial reference frames K or $\mathcal{S}$ can be defined in which Earth's center is immobile.
For quasi-static processes, the $\partial\bm{D}/\partial t$ term in 
$\bm{\nabla}\times\bm{H}$ can also be safely neglected.

For clarity, all discussions in Section \ref{JJderiv} and in Appendix A of the present comment, will use a single frame of reference $\mathcal{S}$ with an origin at Earth's center and inertial reference directions, {\em this frame will be treated as inertial}. For Cartesian or cylindrical coordinates, the z direction will be chosen parallel to the angular velocity $\bm{\omega}=\omega_z\bm{\bm{\hat{z}}}$ of the Earth 
%
\footnote{The standard notation $x,y,z,\rho,\phi$ is used in two different contexts in the present discussion: (a) for the position variables of the inertial frame $\mathcal{S}$ in Section \ref{JJderiv} and Appendix A of the present comment and (b) as position variables in the device bound frame defined in Fig.~1 of Ref.~\onlinecite{Chyba:2016fk}. 
No explicit symbols are assigned for the position variables of frames K in Ref.~\onlinecite{Chyba:2016fk}. This may lead to ambiguity, notably in Eq.~C\&H(26).}. 
%
In the inertial frame $\mathcal{S}$, the Lorentz force $\bm{F}$ acting on a charge $q$ moving at velocity $\bm{v}$ is given by
\jjlabel{geoc06}
\begin{equation}\label{geoc06}
\bm{F}=q(\bm{E}+\bm{v}\times\bm{B}) 
=q\big(-\bm{\nabla}V - \frac{\partial\bm{A}}{\partial t} +\bm{v}\times\bm{B}\big)\,,
\end{equation}
where $\bm{E}$ is the electric field, $V$ is the scalar electric potential, and $\bm{A}$ is the vector potential.
The $\bm{B}$ field in Eq.~(\ref{geoc06}) is the field generated by all sources, including eddy and other currents $\bm{J}$ and magnetization $\bm{M}$ in a ferrite device or other material, and, of course, the Earth bound sources of geomagnetism. 
Eq.~(\ref{geoc06}) will be used 
close to the Earth's crust, 
with $\bm{v}$ 
approximated as the velocity $\bm{\omega}\times\bm{r}$ of the solid conducting medium, 
neglecting the extremely small drift velocity of the mobile charge carriers with respect to this medium.

 The emf in a closed loop, in, through, or around any such passive device, is proportional to the flux of $\bm{\nabla}\times\bm{F}$ through the loop, as 
derived from Eq.~(\ref{geoc06}) by noting that $\bm{\nabla}\times(\partial\bm{A}/\partial t)=\partial\bm{B}/\partial t$ and $\bm{\nabla}\times(\bm{\nabla}V)=0$:
\jjlabel{geoc24}
\begin{eqnarray} 
\frac{1}{q}\bm{\nabla}\times\bm{F}&=&\bm{\nabla}\times\bm{E} 
+ \bm{\nabla}\times[\bm{v}\times\bm{B}]
\nonumber \\  
&=& -\frac{\partial \bm{B}}{\partial t} + \bm{\nabla}\times[\bm{v}\times\bm{B}]\,.
\label{geoc24}
\end{eqnarray}
For a steady rotation at the constant angular velocity 
$\bm{\omega}=\omega_{z} \bm{\hat{z}}$, and for a passive device fixed to the Earth, the relevant electromagnetic quantities rapidly settle to their steady state values $\bm{B}^s$, $\bm{M}^s$, $\bm{E}^s$, $\bm{F}^s$, $\bm{J}^s$, $\cdots$ which are seen as time independent by Earth bound observers, hence all satisfy Eqs.~(\ref{Atest1}) and (\ref{Atest4}), or (\ref{Atest6}) and (\ref{Atest7}). 

In the particular case of the steady state $\bm{B}^s$, Eq.~(\ref{Atest4}) simplifies because $\bm{\nabla}\cdot\bm{B}^s=0$, with the result
\jjlabel{Atest10a}
\begin{equation}\label{Atest10a}
-\frac{\partial\bm{B}^s}{\partial t}+ \bm{\nabla}\times[\bm{v}\times\bm{B}^s]=0,
\end{equation}
valid in the inertial frame $\mathcal{S}$, hence $\bm{\nabla}\times\bm{F}^s=0$,
which implies that no continuous electric power can be generated by any device of the type proposed by Chyba and Hand \cite{Chyba:2016fk}, whatever its size, material, or topology. Obviously, the same conclusion follows from Eq.~C\&H(7) of 
Ref.~\onlinecite{Chyba:2016fk}.
\section{The quantitative calculations in Chyba and Hand's Secs.~IV to IX}
\jjlabel{C+H1}\label{C+H1}
In these calculations, C\&H use three types of reference frame: inertial K frames in which Earth's center is immobile, inertial K' frames in which the center of the ferrite shell is immobile at a specified time, and the non-inertial 
frame in which the ferrite device is immobile.

C\&H's derivation proceeds through changes of reference frame, between inertial and rotating, and often approximates a rotation around a distant axis as a translation. For devices much smaller than Earth's radius, the contribution of these approximations to the predicted power generation will be ignored here, in a first round of discussion, because  
it is negligible compared to that of the confusion of frames described below \cite{Note2}.

In C\&HSec.IV, emf's in closed loops are discussed casually, without clear statement about reference frames or the related contribution $-\partial\bm{B}/\partial t$ to 
$\bm{\nabla}\times\bm{E}$. Nevertheless, C\&H briefly indicate that zero emf's are induced in closed loops rigidly bound to the Earth if 
$\bm{\nabla}\times(\bm{v}\times\bm{B})=0$, where $\bm{v}$ is the loop velocity due to Earth's rotation. This leads the authors 
to propose to violate this requirement with magnetically permeable material in order to recover their hope for nonzero emf in loops bound to Earth. If $\partial\bm{B}/\partial t$ had been properly taken into account, the general conclusion would have emerged that no continuous  production of electric power is possible by the envisioned scheme 
(see Eq.~(\ref{Atest10a})). 
Let me, however, pursue my scrutiny.

In C\&HSec.V, the usual  terms $-\partial\bm{A}/\partial{t}$ or $-\partial\bm{B}/\partial{t}$ reappear in the equations 
and the frames of reference are clearly identified. This leads first 
 to the ``advection-diffusion'' equation for $\bm{A}$ in $K$, Eq.~C\&H(6), which is a combination of the Lorentz force equation 
(Eq.~(\ref{geoc06})) with Maxwell's equations and the phenomenological relations $\bm{J}=(\sigma/q)\bm{F}$ (Ohm's law) and $\bm{B}=\mu\bm{H}$ where $\mu$ is the magnetic permeability of a linear magnetizable material 
\begin{equation*}
\tag*{C\&H(6)}
-\nabla V-\partial\bm{A}/\partial t +\bm{v}\times(\bm{\nabla}\times\bm{A})=\eta\bm{\nabla}\times\bm{\nabla}\times\bm{A},
\end{equation*}
where $\eta=(\sigma\mu)^{-1}$ 
\footnote{The MnZn ferrite (MN60 material) suggested by C\&H manifests conspicuous non-linearity and hysteresis at low fields, hence it is not a linear magnetizable material. This, however, does not affect the final conclusion about continuous electric power generation.},
and $\bm{v}$ is the velocity of the conducting and magnetically permeable material, as measured in frame K.
As indicated by C\&H, the curl of Eq.~C\&H(6) yields the advection-diffusion equation for $\bm{B}$ in K, or ``induction equation'':
\begin{equation*}
\tag*{C\&H(7)}
-\partial\bm{B}/\partial t +\bm{\nabla}\times(\bm{v}\times\bm{B})
  =-(1/\sigma\mu)\bm{\nabla}^2\bm{B}.
\end{equation*}
Unfortunately, C\&H have not noticed that 
Eq.~(\ref{Atest10a}) 
also applies for Eq.~C\&H(7), with the same implication that no electric power can be generated by the device {\em in the steady state}, 
hence, the claimed ``loophole in the proof'' does not exist.
 I shall nevertheless pursue my scrutiny.

In C\&HSec.~VI, the discussion is specialized to a cylindrical shell of 
idealized 
conducting ferrite, for which the calculations can be performed analytically very far by 
extending the results of Prat-Camps  {\em et al.} \cite{Prat-Camps:2012fk} to a device translating at velocity $\bm{v}\ne 0$ in the inertial frame K.
The resulting expressions are approximations valid for $z$ in the central region of a finite length shell.

Beginning in their Sec.~VII, C\&H pursue the discussion together in the device bound frame of Fig.~1 of Ref.~\onlinecite{Chyba:2016fk} (with the corresponding notation for position variables) and, for times very close to $t$, in a K frame 
whose origin coincides with that of the device frame at the exact time $t$.
Using a convenient gauge such that $\bm{\nabla}\cdot\bm{A}=-V/\eta$, C\&H reduce Eq.~C\&H(6) to the single nontrivial equation for $A_z$ in K,
valid in the ferrite device of Fig.~1,
 \begin{equation*}
\tag*{C\&H(25)}
\partial A_z/\partial t+v\,\partial A_z/\partial y=\eta\bm{\nabla}^2A_z\,.
\end{equation*}
%
\footnote{There seems to be a misprint in Ref.~\onlinecite{Chyba:2016fk} on line 11 after Eq. C\&H(25): $v\partial A_z/\partial t$ should be replaced by $v\partial A_z/\partial y$. Similar misprint on line 14.}.
%
Next, C\&H decompose $A_z$ as $A_z=A_s+A_t$, where $A_s$ is the steady state component and $A_t$ the transient component which is expected to decay extremely rapidly. 
For the convenience of the reader, I shall now copy five consecutive lines from C\&H's article:
``The solution to Eq. (25) may, in general, be written as''
\begin{equation*}
\tag*{C\&H(26)}
A_z=A_s(\rho,\phi)+A_t(\rho,\phi,t),
\end{equation*}
``where $A_s(\rho,\phi)$ solves the steady-state equation''
\begin{equation*}
\tag*{C\&H(27)}
v\,\partial A_s/\partial y=\eta\bm{\nabla}^2A_s
\end{equation*}
``and $A_t(\rho,\phi,t)$ solves the time dependent equation $\cdots$''.
The unmotivated absence of a term $\partial A_s/\partial t$ on the l.h.s. of Eq.~C\&H(27) is very surprising: by the definition of the notion of steady state in the present context, $A_s$ is time independent {\em as seen by device (or Earth) bound observers}, hence it has a {\em presumably nonzero} 
partial time derivative as seen from the K frame in which Eqs.~C\&H(6, 7, 25, 27) are valid.
The missing term is easily evaluated as $\partial A_s/\partial t=-v\partial A_s/\partial y$ 
(in the usual C\&H approximation of describing exact rotation by a translation), with the conclusion that $\bm{\nabla}^2 A_s =0$, hence zero electric current circulates in the device in the steady state. This conclusion remains valid if Earth's rotation is taken into account {\em exactly}, in agreement with widespread expectation and with my own prediction in Sec.~II above.

The quantitative discussion presented in C\&HSec.XI ``Analysis in the laboratory frame" takes for granted the erroneous deduction of Eq.~C\&H(27) from Eq.~C\&H(25), hence it does not provide a valid confirmation of the previous prediction in frame K.

It is tempting to conjecture that the unfortunate absence of the $\partial A_s/\partial t$ term on the l.h.s. of  Eq.~C\&H(27) is due to the visually misleading use, in Eq.~C\&H(26), of position variables $\rho$ and $\phi$ which belong to the non-inertial device-centered frame, whereas the relevant equations of motion are valid in the inertial frame K.
Whatever the mechanism of the confusion, the absence of steady-state power generation  makes Ref.~\onlinecite{Chyba:2016fk} almost pointless. 
 It also makes dubious a similar prediction by Hand et al. \cite{Hand:2011fk}.

%
\section{Other questionable aspects of Chyba and Hand's article}
\jjlabel{questionable}\label{questionable}
\subsection{Intrinsic velocity of quasi-static magnetic fields, axisymmetric and non-axisymmetric components}
\jjlabel{intrinsic}\label{intrinsic}
The notion that quasi-static magnetic fields $\bm{B}$ have an intrinsic velocity (somewhat like ordinary massive particles or continuous media) was originally introduced among attempts 
to reconcile previous conventional wisdom with new theories and experiments 
during the long historical controversy about the validity of special relativity, the interpretation of Maxwell's equations, and the existence of aether as a support of electromagnetic phenomena. After the abandonment of the aether hypothesis, it became doubtful that a notion of intrinsic velocity for $\bm{B}$ fields could be introduced that would be strictly compatible with Maxwell's equations and special relativity, even in the weakly relativistic approximation. Recent publications on this topic (e.g. Galili and Kaplan \cite{Galili:1997fk}) strongly advise to systematically avoid  
 the notion of velocity of a $\bm{B}$ field, and recommend to follow the example of the  
 textbooks (e.g. Landau and Lifshitz \cite{Landau:1984fk,Landau:1971fk}, Reitz and Milford \cite{Reitz:1960fk} or Jackson \cite{Jackson:1962fk}) which do that.
Later, a related notion of velocity has been successfully introduced 
to guide intuition 
in the very different context of magnetohydrodynamics ( 
see e.g.  Spruit \cite{Spruit:fk}).

Unexpectedly, the legacy notion that quasi-static $\bm{B}$ fields have an intrinsic velocity pervades C\&H's article as a background concept for ``intuitive'' discussions, 
further complicated by the unwarrantable idea that, in problems dominated by axial rotation, the axisymmetric and non-axisymmetric parts of the $\bm{B}$ field may have different intrinsic velocities (see last paragraph of page 14 of Ref.~\onlinecite{Chyba:2016fk}), and that only the axisymmetric part contributes to electromagnetic induction and Lorentz force.

Many ``intuitive'' discussions hinge on the dimensionless ``magnetic Reynolds number'' $R_m$ which, surprisingly, is never mentioned in the standard derivation \cite{Note1} of the related display equations.

In C\&HSec.~X ``Intuitive Physical Picture'', the authors describe an extremely strange ``intuitive'' model in which its
zero
 intrinsic velocity forces the $\bm{B}$ field to penetrate into the moving ferrite device, where it is subsequently relaxed to its equilibrium value.

In C\&HSec.~XI, in a further promotion of the 
assumed 
existence of an intrinsic velocity of quasi static $\bm{B}$ fields, the authors claim that clear conclusions concerning this velocity can be deduced from highly respected experimental results that are one century old (C\&H references [1], [15], and [19]), although these experimental results are also in perfect agreement with predictions from today's electromagnetism which ignores and rejects the notion of intrinsic velocity of $\bm{B}$ fields.
\subsection{Observation and use of the predicted e.m.f.'s}
\jjlabel{SsecObs}\label{SsecObs}
The elaborate discussion of the electromagnetic state of the {\em isolated} ferrite device presented in C\&HSec.~VII is in surprising contrast with the very naive ideas 
 (reminiscent of Chyba at al. \cite{Chyba:2015fk}) 
 invoked to treat the steady state of the {\em connected} device in the perspectives of measurements of the predicted emf's or of practical use of the generated electric power. 
For instance, the statement that ``half the emf would be measured across the (d, f) diagonal'' is justified or explained neither in the caption of Fig. 1 nor in the paragraph around Eq. C\&H(88).
\appendix
\section{Inertial evolution of vector and scalar fields bound to the rotating Earth}
In the steady state of the present model, many relevant vector and scalar fields are time independent {\em as seen by Earth bound observers}.
Let $\bm{G}^s(\bm{r};t)$ describe such a vector field depending on position 
$\bm{r}=x\hat{\bm{x}}+y\hat{\bm{y}}+z\hat{\bm{z}}$ in the inertial frame ${\cal S}$, and on time $t$. 
Let also Earth's rotation in frame ${\cal S}$ be described by the
 rotation operator ${\cal R}_z(\omega_z\tau)$ which rotates vectors by an angle $\omega_z\tau$ around the $z$ axis of frame ${\cal S}$, for instance, 
\jjlabel{Atest2}
\begin{eqnarray} &{}&
{\cal R}_z(\omega_z\tau)\bm{\hat{x}}=
\bm{\hat{x}}\cos[\omega_z\tau]+\bm{\hat{y}}\sin[\omega_z\tau]\,,
\nonumber \\  &{}&
{\cal R}_z(\omega_z\tau)\bm{\hat{y}}=
\bm{\hat{y}}\cos[\omega_z\tau]-\bm{\hat{x}}\sin[\omega_z\tau]\,,
\hspace{7pt} {\cal R}_z(\omega_z\tau)\bm{\hat{z}}= \bm{\hat{z}}\,,
\nonumber \\ &{}&
{\cal R}_z(\omega_z\tau)\bm{r}=\bm{\hat{x}}(x\cos[\omega_z\tau]-y\sin[\omega_z\tau])
\nonumber \\ &{}& \hspace{47pt}
+\,\bm{\hat{y}}(y\cos[\omega_z\tau]+x\sin[\omega_z\tau])+\bm{\hat{z}}z \,.
\label{Atest2}
\end{eqnarray}
 If the vector field $\bm{G}^s(\bm{r};t)$ is immobile as seen by Earth bound observers, 
then its projections on $\bm{\hat{x}}$, $\bm{\hat{y}}$, $\bm{\hat{z}}$ at time $t$, 
and on 
$[{\cal R}_z(\omega_z\tau) \bm{\hat{x}}]$, $[{\cal R}_z(\omega_z\tau) \bm{\hat{y}}]$,
$\bm{\hat{z}}$ at time $t+\tau$ satisfy
%
\jjlabel{Atest1}
\begin{eqnarray}
\bm{\hat{x}}\cdot\bm{G}^s(\bm{r};t)&=&[{\cal R}_z(\omega_z\tau) \bm{\hat{x}}]\cdot
\bm{G}^s({\cal R}_z(\omega_z\tau)\bm{r};t+\tau),
\nonumber \\
\bm{\hat{y}}\cdot\bm{G}^s(\bm{r};t)&=&[{\cal R}_z(\omega_z\tau) \bm{\hat{y}}]\cdot
\bm{G}^s({\cal R}_z(\omega_z\tau)\bm{r};t+\tau),
\nonumber \\
\bm{\hat{z}}\cdot\bm{G}^s(\bm{r};t)&=& \label{Atest1}
G_z^s({\cal R}_z(\omega_z\tau)\bm{r};t+\tau),
\end{eqnarray}
and the corresponding partial time derivative of $\bm{G}^s(\bm{r};t)$ in the inertial frame ${\cal S}$ is easily evaluated from the systematic expansion of Eqs.~(\ref{Atest1}) in power series of $\tau$ (limited to first order). For instance, the first line of Eqs.~(\ref{Atest1}) gives, successively, with due reference to Eqs.~(\ref{Atest2}),
\jjlabel{Atest3}
\begin{eqnarray} &{}&
G_x^s(\bm{r};t)
\nonumber \\ &{}& \hspace{10pt}
=[\bm{\hat{x}}+\omega_z\tau\bm{\hat{y}}+\cdots]\cdot
\bm{G}^s\big[\bm{\hat{x}}(x-\omega_z\tau y+\cdots)
\nonumber \\ &{}& \hspace{86pt}
+\bm{\hat{y}}(y+\omega_z\tau x+\cdots)+\bm{\hat{z}}z;t+\tau \big]
\nonumber \\ &{}& \hspace{10pt}
=G_x^s(\bm{r};t)+\omega_z\tau G_y^s(\bm{r};t)
\nonumber \\ &{}& \hspace{30pt}
+\,\omega_z\tau\left(-y\frac{\partial G_x^s}{\partial x}+x\frac{\partial G_x^s}{\partial y}  \right)
+\tau\frac{\partial G_x^s}{\partial t} +\cdots \,,
\label{Atest3}
\end{eqnarray}
%
%
%
%
and the complete result can be written as 
%
\jjlabel{Atest4m}
\begin{eqnarray} &{}&
\frac{\partial\bm{G}^s}{\partial t}=\omega_z\bigg\{ \bm{\hat{x}} \left(+y\frac{\partial G_x^s}{\partial x}
-x\frac{\partial G_x^s}{\partial y}-G_y^s\right) +
\nonumber \\ &{}& \hspace{50pt}
+\bm{\hat{y}} \left( +y\frac{\partial G_y^s}{\partial x}
-x\frac{\partial G_y^s}{\partial y}+G_x^s\right) +
\nonumber \\ &{}& \hspace{50pt}
+\bm{\hat{z}} \left( +y\frac{\partial G_z^s}{\partial x}
-x\frac{\partial G_z^s}{\partial y}\right) \!\! \bigg\},
\label{Atest4m}
\end{eqnarray}
where all quantities in Eqs.~(\ref{Atest4m}) are evaluated at $(\bm{r};t)$. 
The terms $-G_y^s$ and $+G_x^s$ in Eqs.~(\ref{Atest4m}) arise from the rotation of the Earth bound reference directions (see Eqs.~(\ref{Atest2}, \ref{Atest1})).
With $\bm{v}=(\omega_z \bm{\hat{z}})\times \bm{r}$, Eq.~(\ref{Atest4m}) can be written under the more convenient compact form
\jjlabel{Atest4}
\begin{equation}\label{Atest4}
\frac{\partial\bm{G}^s}{\partial t}=
\bm{\nabla}\times(\bm{v}\times\bm{G}^s) + \bm{v}\,(\bm\nabla\cdot\bm{G}^s).
\end{equation}

Let now ${\cal G}^s(x,y,z;t)$ describe a {\em scalar field} that is time independent as seen by Earth bound observers:
\jjlabel{Atest6,7}
\begin{eqnarray} &{}&
{\cal G}^s(x,y,z;t)={\cal G}^s\big(x\,\cos[\omega_z \tau]+y\,\sin[\omega_z \tau],
\nonumber \\ \label{Atest6}&{}& \hspace{70pt}
 y\,\cos[\omega_z \tau]-x\,\sin[\omega_z \tau],z;t+\tau\big),\;\;\;\;
 \\ &{}& \text{hence,}\hspace{10pt}
 \frac{\partial{\cal G}^s}{\partial t}=-\omega_z\Big(y\frac{\partial{\cal G}^s}{\partial x}
   -x\frac{\partial{\cal G}^s}{\partial y} \Big) 
   =\bm{v} \cdot (\bm{\nabla} \mathcal{G}^s), \hspace{18pt}
\label{Atest7}
\end{eqnarray}
where all quantities in Eq.~(\ref{Atest7}) are evaluated at $(x,y,z;t)$.
%
%
%
\end{document}